\documentstyle[psfig,epsf]{article}
\begin{document}

\begin{center}
{\Large \bf Automation of  Feynman diagram evaluation} \\

\vspace{4mm}
M.~TENTYUKOV\footnote{
Supported by Bundesministerium f\"ur Forschung und Technologie
under PH/05-7BI92P 9.
}\\
Joint Institute for Nuclear Research,\\
141980 Dubna, Moscow Region, Russian Federation.\\ E-mail:
tentukov@thsun1.jinr.dubna.su

\vspace{2mm}

J.~FLEISCHER\\
Fakult\"at f\"ur Physik, Universit\"at Bielefeld\\
D-33615 Bielefeld, Germany\\
~E-mail: fleischer@physik.uni-bielefeld.de.\\

\begin{abstract}
A C-program
DIANA (DIagram ANAlyser) for the automation of Feynman diagram 
evaluations is presented. 
\end{abstract}

\end{center}

   Recent high precision experiments require, on the side of the theory,
high-precision calculations resulting in the evaluation of higher
loop dia\-grams in the Standart Model (SM).  
For specific processes thousands of multiloop Feynman
dia\-grams do contribute, and it turns out to be impossible to perform 
these calculations by hand. This makes the request for automation a
high-priority task. 

Several different packages
have been developed with different areas of applicability.
For example,
FEYNARTS / FEYNCALC \cite{FeynmArts} are MATHEMA\-TICA packages convenient
for various aspects of the calculation of radiative corrections in the SM.
There are several FORM packages for evaluating multiloop diagrams, like
MINCER \cite{MINCER}, and a package \cite{leo96} 
for the calculation of 3-loop bubble integrals with one non-zero mass.
Other packages for automation are
GRACE \cite{GRACE} and COMHPEP \cite{CompHep},
which partially perform full calculations, from the process definition 
to the cross-section values. 

A somewhat different approach is persued  by
XLOOPS \cite{XLoops}. 
A graphical user interface
makes XLOOPS an `easy-to-handle' program package, but is mainly aimed 
to the evaluation of single diagrams.
To deal with thousands of diagrams, it is neccessary to use 
special techniques like databases and special controlling programs.
In \cite{Vermaseren} for evalua\-ting more than 11000 diagrams
the special database-like program MINOS was developed.
It calls the relevant FORM programs, waits until they fi\-nished, picks 
up their results and repeats the process without any human interference.

It seems impossible to develop an universal package, which 
will be effective for all tasks. It appears
absolutely necessary that various groups produce their own solutions
of handling the problem of automation:
various ways will be of different
efficiency, have different domains of applicability, and last but not
least, should eventually allow for completely independent checks of
the final results.
This point of view motivated us to seek our own
way of automatic evaluation of Feynman diagrams. 

Our first step is dedicated to the automation of 
the muons two-loop anomalous magnetic moment (AMM) 
${\frac{1}{2}(g-2)}_{\mu}$. For this purpose
the package TLAMM was developed \cite{TLAMM}.
The algorithm is implemented
as a FORM-based program package. For generating and automatically
evaluating any number of two-loop self-energy diagrams, a special
C-program has been written. This program creates the initial
FORM-expression for every diagram generated by 
QGRAF \cite{QGRAF}, executes the
corresponding subroutines and sums up the various contributions.
In the SM 1832 two-loop diagrams contribute in this case. 
The calculation of the bare diagrams is finished.

Our aim is to create some universal
software tool for piloting the process of generating the source
code in multi-loop order
for analytical or numerical evaluations and to keep the control of
the process in general. Based on this instrument, we can attempt to build 
a complete package performing the computation of any given process, at least
in the framework of a concrete model. 

The project called DIANA (DIagram ANAlyser) \cite{FT} for the
evalua\-tion of Feynman diagrams is being finished by our group at
present.

The program DIANA  contains two ingredients:
1.~Analyzer of diagrams.
2.~Interpreter of a special text manipulating language (TM).
The TM language is a very simple TeX-like language for creating
source code and organizing the interactive dialog.

The analyzer reads QGRAF output and passes necessary
information to the interpreter. For each diagram the interpreter performs
the TM-program, producing input for further evaluation of the diagram.
Thus the program:

Reads QGRAF output and for each diagram it:
1.~Determines the topology,
looking for it in the table of
all known topologies and distributes momenta according to the current topology.
If we do not yet know all needed topologies,
we may use the program
to determine missing topologies that occur in the process.
2.~Creates an internal representation of the diagram 
in terms of vertices and propagators
corresponding to the Feynman integrand.
3.~Invokes the interpreter to execute the TM-program 
to insert explicitely expressions for the vertices, propagators etc.
Producing FORM input e.g., is done by passing to it the necessary
information from the analyser.

Executing the TM-program provides apart from the possibility  to
calculate each
diagram using FORM or another formulae manipulating language, to do 
some numerical
calculation by means of FORTRAN, to create
a postscript file for the  picture of the current diagram, etc.\\

The program operates as follows:
first of all, it reads its configuration 
file, which may be produced manually or by DIANA as well. This file contains:
1.~The information about various settings (file names, numbers
of external particles, definition of key words, etc.)
2.~Momenta distribution for each topology.
3.~Description of the model (i.e., all particles, propagators and vertices).
4.~TM-program.
Then the program starts to read QGRAF output.
For each diagram it determines the topology,
assigns indices and creates the textual representation
of the diagram corresponding to the Feynman integrand. All defined data
(masses of particles, momenta on each lines, etc.) are stored
in internal tables, and may be called by TM-program operators.
At this point DIANA invokes the interpreter and performs the TM-program.
After that it starts to work with the next diagram.

When all diagrams are processed, the program may call the interpreter
once again (optionally) to perform the TM-program a last time. Such last call
may be used to do some final operations like summing up the
results.

There is a possibility to use the program as an interpreter only.
If one specifies in the configuration file
\begin{verbatim}
only interpret
\end{verbatim}
then DIANA will not try to read QGRAF output, but immediately
enters the TM - program.\\

\begin{figure}[ht]
\centerline{\vbox{\epsfysize=40mm \epsfbox{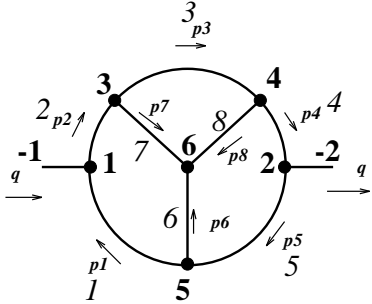}}}
\caption{\label{topol1} Topology
(-2,2)(-1,1)(5,1)(1,3)(3,4)(4,2)(2,5)(5,6)(3,6)(4,6).
    }
\end{figure}

Topologies are represented in terms of
ordered pairs of numbers like (fromvertex,tovertex) (see fig.~\ref{topol1}).
All external legs have negative numbers. These are supposed to begin with -1. 
The fist numbers correspond to ingoing particles, the last numbers to outgoing
ones. External legs must be connected with vertices of smallest possible 
identifying number. 
The number of an internal line corresponds to its position in the
chain of pairs and the direction from the first to the second number
in the pair:
$$
\begin{array}{ccccccccc}
\mbox{direction:}&5\to 1&1\to 3&3\to 4&4\to 2&2\to 5&5\to 6&3\to 6&4\to 6\\
(-2,2)(-1,1)&(5,1)&(1,3)&(3,4)&(4,2)&(2,5)&(5,6)&(3,6)&(4,6)\\
\mbox{number:}&1&2&3&4&5&6&7&8
\end{array}
$$
Knowing thus the topology we can assign momenta.
Their distribution according to fig. \ref{topol1} e.g. is added like 
\begin{verbatim}
topology = (-2,2)(-1,1)(5,1)(1,3)(3,4)(4,2)(2,5)(5,6)(3,6)(4,6):
  p1,p2,p3,p4,p5,p6,p7,p8;
\end{verbatim}
This fixes directions and values of all momenta on internal lines.

The main goal of the TM-language is the creation of text files. 
In principle we could have used one of the existing languages, 
but we want a
very specific language: it should be powerful enough to create 
arbitrary program texts. On the other hand, it should be 
very simple and easy-in-use, so that even non-programmers can use it.

Similar to the TeX language, all lines without special escape - 
characters (``$\backslash$'') are simply typed to the output file. 
So, to type
``Hello, world!'' in the file ``hello'' we may write down the following
program:

\begin{verbatim}
\program
\setout(hello)
Hello, world!
\end{verbatim}

Each word
the first character of which is the escape character will be considered as a command.
This feature makes this language very easy-to-use.

The user types his FORM program and has the possibility to insert 
in the same line TM-operators as well.
For example, the typical part of a TM program looks like follows:
\begin{verbatim}
\setout(d\currentdiagramnumber().frm)
#define dia "\currentdiagramnumber()"
#define TYPE "\type()"
#define COLOR "\color()"
#define LINES "\numberofinternallines()"
\masses()
#include def.h
l  R=\integrand()
#call feynmanrules{}
#call projection{}
#call reducing{'TYPE'}
#call table{'TYPE'}
#call colorfactor{'COLOR'}
.sort
drop R;
g dia'dia' = R;
.store
save dia'dia'.sto;
.end
\setout(null)
\system(\(form -l )d\currentdiagramnumber().frm)
\end{verbatim}

Some of the TM - commands are just TM-operators while some are functions 
(returning a value) written in 
the TM-language itself. For example, \verb|\numberofinternallines()| is 
a built-in TM operator, while the function \verb|\masses()| is the 
TM-language function:
\begin{verbatim}
\function masses;\-\let(i,0)
\+\do
#define m\inc(i,1) "\mass(\get(i))"
\while "\numcmp(\get(i),\numberofinternallines())" eq "<" loop
\end
\end{verbatim}

This TM-program will generate the FORM input for each diagram and 
then it will call the FORM interpreter by means of the operator 
\verb|\system()|. For example, the corresponding part of 
the FORM program generated for diagram number 15 looks like follows:
\begin{verbatim}
#define dia "15"
#define TYPE "4"
#define COLOR "3"
#define LINES "4"
#define m1 "mmH"
#define m2 "mmW"
#define m3 "mmW"
#define m4 "mmH"
#include def.h
l  R=
       1*V(1,mu1,mu,2)*(-i_)*em^2/2/s*V(2,0)*(-i_)*1/4*em^2/s^2*mmH/mmW*
       V(3,mu2,+q4-(+q3),1)*(-i_)*em/2/s*SS(1,0)*i_*VV(2,mu1,mu2,+q2,2)*i_*
       SS(3,2)*i_*SS(4,0)*i_;
#call feynmanrules{}
#call projection{}
#call reducing{'TYPE'}
#call table{'TYPE'}
#call colorfactor{'COLOR'}
.sort
drop R;
g dia'dia' = R;
.store
save dia'dia'.sto;
.end
\end{verbatim}

At present, we have finished the C-part of this project.
Also, we have several files with TM-language macros to start DIANA.
Similar as LaTeX provides the possibility for non-specialsts to
typeset high-quality texts using TeX language, 
these macros permit DIANA to work at very high level. The user
can specify the model and the process, and DIANA will generate all neseccary 
files. 

\vskip 5mm
\noindent
{\noindent\bf Acknowledgements}
\vglue 0.2cm
M.T. acknowledges the University of Bielefeld for
the warm hospitality.

This paper was supported in part by RFFI grant \# 96-02-17531.


\begin{thebibliography}{99}

\bibitem{FeynmArts}
J.~K\"ublbeck, M.~B\"ohm and A.~Denner,
Comp. Phys. Comm. 60 (1990) 165;
~~R.~Mertig, M.~B\"ohm and A.~Denner,
Comp. Phys. Comm. 64 (1991) 345.
\vspace{-2.5mm}
\bibitem{MINCER}
S.A.~Larin, F.V.~Tkachov, J.A.M.~Vermaseren, NIKHEF-H/91-18.
\vspace{-2.5mm}
\bibitem{leo96}
L.V.~Avdeev, Comp. Phys. Comm. {\bf 98} (1996) 15.
 \vspace{-2.5mm}
\bibitem{GRACE}
T.~Ishikawa {\it et al}, Minami-Taeya group ``GRACE manual'', KEK-92-19, 1993.
\vspace{-2.5mm}
\bibitem{CompHep}
E.E.~Boos {\it et al}, SNUTP 94-116 (1994); (hep-ph/9503280).
\vspace{-2.5mm}
\bibitem{XLoops}
L.~Br\"ucher {\it et al}, Nucl.Instrum. Meth. {\bf A389} (1997) 323;
~~A.~Frink, J.G.~K\"orner and J.B.~Tausk, (hep-ph/9709490).
\vspace{-2.5mm}
\bibitem{Vermaseren}
T.~van Ritbergen et al., Int. J. Mod. Phys. {\bf C6} (1995) 513.
\vspace{-2.5mm}
\bibitem{TLAMM}
L.V.~Avdeev, J.~Fleischer, M.~Yu.~Kalmykov, M.~Tentyukov,  
Nucl.Instrum. Meth. A 389 (1997) 343;
{\it Towards Automatic Analytic Evaluation of Diagrams with Masses},
accepted for publication in Comp.Phys.Comm., (hep-ph/9710222);
L.V.~Avdeev, M.Yu.~Kalmykov, Nucl. Phys. {\bf B502} (1997) 419.
\vspace{-2.5mm}
\bibitem{FT}
J. Fleischer and M. Tentukov, {\it A Feynman Diagram Analyser DIANA},
Bielefeld preprint \mbox{BI-TP-97/44}, in preparation.
\vspace{-2.5mm}
\bibitem{QGRAF}
P.~Nogueira, J. Comput. Phys. {\bf 105} (1993), 279.
\vspace{-2.5mm}
\end{thebibliography}
\end{document}